\documentclass[prd,aps,nofootinbib,nobibnotes,notitlepage,superscriptaddress,twocolumn]{revtex4-1}
\usepackage{appendix}
\usepackage{comment}
\usepackage{color}
\usepackage{amsmath}
\usepackage{amsfonts}
\usepackage{graphicx}
\usepackage[dvipsnames]{xcolor}
 \usepackage[colorlinks=true,hyperfootnotes=true,citecolor=cyan]{hyperref}
\usepackage[utf8]{inputenc}
\usepackage{float}

\newcommand{\rs}{r_{\rm s}}
\newcommand{\as}{a_{\rm s}}

\newcommand{\dd}{{\rm d}}

\newcommand{\Lag}{\mathcal{L}}

\newcommand{\mU}{\mathcal{U}}

\newcommand{\pred}{p_d}
\newcommand{\be}{\begin{equation}}
\newcommand{\ee}{\end{equation}}
\newcommand{\bea}{\begin{eqnarray}}
\newcommand{\eea}{\end{eqnarray}}

\newcommand{\mpl}{M_{\rm Pl}}
\newcommand{\mK}{{\mathcal{K}}}

\newcommand{\Lambdae}{\Lambda_{\rm BI}}

\begin{document}

\title{Inhomogeneous Hubble diagram from vector K-mouflage}

\author{Jose Beltr\'an Jim\'enez}
\email[]{jose.beltran@usal.es}
\affiliation{Departamento de F\'isica Fundamental and IUFFyM, Universidad de Salamanca, E-37008 Salamanca, Spain.}
\author{Dario Bettoni}
\email[]{bettoni@usal.es}
\affiliation{Departamento de F\'isica Fundamental and IUFFyM, Universidad de Salamanca, E-37008 Salamanca, Spain.}

\author{Philippe Brax}
\email[]{philippe.brax@ipht.fr}
\affiliation{Institut de Physique  Th\'eorique, Universit\'e Paris-Saclay,CEA, CNRS, F-91191 Gif-sur-Yvette Cedex, France.}
\begin{abstract}
In this Letter we construct the Hubble diagram for a Universe where dark matter is universally charged under a dark non-linear electromagnetic force which features a screening mechanism of the K-mouflage type for repulsive forces. By resorting to the Newtonian approximation, we explicitly show that the cosmological evolution generates an inhomogeneous Hubble diagram that corresponds to a curvature dominated expansion at short distances and converges to the cosmological one of $\Lambda$CDM. We discuss the potential impact of this inhomogeneous profile on the Hubble tension. For completeness, we explicitly show how the Newtonian approximation can be derived from an inhomogeneous relativistic Lema\^itre model.
\end{abstract}

\maketitle

\section{Introduction}
\label{sec:intro}

One of the pressing questions in contemporary cosmology is the apparent discrepancy in the measured value of the Hubble-Lema\^itre constant $H_0$ as deduced from local observations \cite{Riess:2019cxk} and the early time universe \cite{Aghanim:2018eyx} (see also \cite{Bernal:2016gxb,Verde:2019ivm} and references therein). This tension has triggered the search for potential explanations with various mechanisms being advocated~\cite{Poulin:2018cxd,Mortsell:2018mfj,Knox:2019rjx,Alexander:2019rsc,Lin:2019qug,Zumalacarregui:2020cjh,Ballesteros:2020sik,Braglia:2020iik,Desmond:2020wep}. In this Letter, we will elaborate further on the potential explanation for this mismatch along the lines of \cite{Jimenez:2020bgw,BeltranJimenez:2020csl}. In this approach we consider  dark matter to be charged under a dark, non-linear electromagnetic interaction. Then, if a net dark charge is present a repulsive force will be generated between dark matter particles that hence competes with gravity. Models in which a net  overall charge in the Universe is present have a long history and have been explored in the case of both electric \cite{Lyttleton:1959zz,Ignatiev:1978xj} and dark charges either preserving \cite{Gradwohl:1992ue,Kaloper:2009nc,Jimenez:2020bgw,BeltranJimenez:2020csl} or breaking \cite{1979ApJ...227....1B} gauge invariance;  possible mechanisms to produce the separation or asymmetry between particles and anti-particles have also been proposed, see \cite{Goolsby-Cole:2015chd} and references therein. 

In order to get a deeper physical intuition of the cosmological evolution in this class of models, it is convenient to foliate the Universe 
by expanding spherical shells.\footnote{This Newtonian approach to cosmology correctly reproduces the relativistic dynamics in the appropriate limits \cite{Gibbons:2013msa,HARRISON1965437}.} In the early universe, thanks to the non-linear nature of the electromagnetic sector that provides a natural screening mechanism \`a la K-mouflage \cite{Brax:2012jr}, the electric force is absent and the cosmological evolution is oblivious to the presence of charges. Hence,  the motion is co-moving, i.e., all the shells experience the same expansion with the same scale-factor. However, as we will see, the shells of smaller radii gradually exit their screening radii and start experiencing the repulsive electromagnetic interaction. As a result the Universe becomes inhomogeneous, albeit still isotropic, and it can be described relativistically by a Lema\^itre model.\footnote{In \cite{Kaloper:2009nc} a relativistic extension of a model in which a fraction of dark matter is charged is obtained in terms of McVittie solutions \cite{McVittie:1933zz} when the charge equals the mass \cite{Kastor:1992nn}.}

The inhomogeneity of the Universe has important consequences for the local Hubble rate as measured by an observer receiving light signals from an emitter. When both the emitters and the observer are within their screening radii, they both follow the cosmological Hubble flow felt on large scales and the local and cosmological Hubble rates coincide. On the other hand, when both the observer and the emitter have exited their screening radii, the local and cosmological Hubble rates still coincide although in this regime the universe is heavily influenced by the negative curvature induced by the repulsive electromagnetic interaction. Finally, a more interesting situation occurs when the emitters are still screened and the observer has exited its screening radius. In this case, the local Hubble rate has two relevant features. First of all, the local rate is larger than the cosmological Hubble rate for distances between the emitters and the observer large enough compared to the distance of the observer to the centre of the Universe (or of the local and inhomogeneous patch of the Universe to which the observer belongs). Second, the Hubble rate is highly anisotropic, unless the observer is at the centre of the Universe, and such an anisotropy could appear as a scatter in the Hubble diagram.\footnote{It is worth stressing that there are no reasons a priory that requires that the observer should be close to the centre.}

The paper is organized as follows. In section \ref{sec:model}, we define the models of dark electrodynamics and the associated screening phenomenon. We have used the Born-Infeld theory as an example of non-linear electrodynamics and used it for our numerical calculations although the results are general as long as there is no shell crossing. We also describe the Newtonian dynamics of the universe when dark matter is subject to the dark electromagnetic interaction.  The local Hubble diagram and the local Hubble rate in the inhomogeneous Universe resulting from the presence of screened electrodynamics is presented in section \ref{sec:hubble}. We then discuss our conclusions in section \ref{sec:conc}. The derivation of the Newtonian approximation from the relativistic dynamics is given in appendix~\ref{sec:equations} while  in appendix \ref{app:DSE}  we discuss the description of the dark matter in terms of a complex scalar field and its link with a charged fluid subject to dark electromagnetism.

\section{The model}
\label{sec:model}

Let us start by introducing the model that we will consider for the interaction between dark matter particles. We will use the paradigmatic example of Born-Infeld (BI) electromagnetism \cite{Born:1933pep,Born:1934gh} described by the following Lagrangian
\be
\Lag=\Lambdae^4\left[1-\sqrt{-\det\left(g_{\mu\nu}+\frac{1}{\Lambdae^4} F_{\mu\nu}\right)}\right]+ J^\mu A_\mu\,,
\ee
where $\Lambdae$ is the Born-Infeld scale determining when the non-linearities become relevant, $F_{\mu\nu}=\partial_\mu A_\nu-\partial_\nu A_\mu$ the field strength and $J^\mu$ describes the coupling to the conserved current of dark matter. Although we will use this particular model, as we will see, the key feature is the presence of a screening scale, which is a common property of general non-linear electromagnetism so  our results can be straightforwardly extended to the case of more general non-linear electromagnetism. Let us then review how this screening mechanism works for the BI model \cite{Boillat:1970gw}. For that, we consider a spherically symmetric and static lump of dark matter that will then generate a radial electric field given by 
\be
E=\frac{1}{\sqrt{1+\left(\frac{\rs}{r}\right)^4}}\frac{Q}{4\pi r^2}\,,
\ee
with $Q$ the total electric charge of the lump and we have introduced the screening radius
\be
\rs\equiv\sqrt{\frac{Q}{4\pi}}r_{\rm BI}\,,
\ee
where we have defined $r_{\rm BI}=\Lambdae^{-1}$. 
At large distances $r\gg\rs$, the electric field reduces to the usual Maxwellian one $E\simeq\frac{Q}{4\pi r^2}$, while at short distances $r\ll\rs$ the electric field saturates to the value $E\simeq \Lambdae^2$. This means that two lumps of dark matter separated by a distance larger than the sum of their corresponding screening radii will be repelled by a Maxwellian force, while if they are inside the screening region the force is strongly suppressed. 

This is the standard screening mechanism of these models. In the following we will show how this phenomenology gives rise to inhomogeneous cosmologies. For that it will be sufficient to use the Newtonian approximation, although a general relativistic formulation in terms of Lema\^itre models exist (see \cite{BeltranJimenez:2020csl} and Appendix \ref{sec:equations}). Thus, let us consider a Universe filled with a homogeneous and isotropic distribution of dark matter particles, all of which are universally coupled under the BI field. The interaction between the particles will then be the gravitational attraction plus the electric repulsion. If we divide our distribution of dark matter into concentric spherical shells, the evolution of the shell of radius $R$ will be governed by the equation
\be
\ddot{R}=-\frac{GM(R)}{R^2}\left[1-\frac{\beta^2}{\sqrt{1+\left(\frac{\rs(R)}{R}\right)^4}}\right]\,,
\label{eq:eqR}
\ee
where $G$ is Newton's constant, $M(R)$ is the mass contained within the shell and $\beta$ is the charge-to-mass ratio of the dark matter particles defined as $\beta\equiv \sqrt{2}q\mpl/m$ which we expect to be of order one and where the screening radius depends on $R$ as it is determined by the total charge enclosed by the corresponding sphere. It is worth noticing that the crucial minus sign in front of $\beta^2$ leading to repulsive forces between alike charges originates from the spin-1 nature of the gauge field. This is in high contrast with the more common case of scalar $K-$mouflage where the additional force is attractive and, therefore, the phenomenology that we will obtain here is not possible for the scalar counterpart.  

If the shells do not cross in the evolution (as it is the case of BI electromagnetism) then the mass within each shell is conserved and, if the initial density profile is homogeneous, we can express it as
\be\label{eq:mass_Newt_d}
M=\frac{4\pi}{3}\rho_\star r^3\,,
\ee
where $\rho_\star$ is the constant density and we have introduced the Lagrangian coordinate $r$ given by the radius of the shell at some initial time $t_\star$, i.e., $r=R(t_\star)$. For the same reasons, the total charge within each shell is conserved throughout the evolution and we will have $Q\propto r^3$ so that the screening scale does not depend on time and it will be given by 
\be
\rs=\sqrt{\frac{\beta M}{4\sqrt{2}\pi\mpl}}r_{\rm BI}=\sqrt{\frac{\beta \rho_\star}{3\sqrt{2}\mpl}}r^{3/2}r_{\rm BI}\,,
\ee
so larger shells have larger screening radii. {From this expression we can also verify that the region below $\rs$ where classical non-linearities are important is parameterically above the quantum strong coupling scale given by $r_{\rm BI}$. Assuming $\beta\lesssim \mathcal{O}(1)$, this is the case for astrophysical objects which have masses $M\gg \mpl$. \footnote{Out of curiosity, this occurs for objects containing more than $\sim10^{-5}N_A$ nucleons, with $N_A$ the Avogadro's number.}} If we introduce the scale factor $a(t,r)\equiv R(t)/r$, the equation can be written as 
\be
\ddot{a}=-\frac{4\pi G \rho_\star}{3a^2}\left[1-\frac{\beta^2}{\sqrt{1+\left(\frac{\as}{a}\right)^4}}\right]\,,
\label{eq:eqa}
\ee
with $\as=\rs/r$. We can see that the whole dependence on $r$ is contained within the non-linearities of the BI force. Thus, at early times when the BI force is negligible, the system evolves in a comoving motion so the initial homogeneity and isotropy is preserved. However, as the different shells cross their screening radius, they start evolving differently and, as a consequence, an explicit dependence on $r$ appears. Asymptotically, when all the relevant shells have exited their screening radius, the force is again of the Coulombian type for all the shells so the motion becomes $r$-independent again. However, there is an induced $r$-dependence originating from the screening scale $\rs(r)$ that is maintained in the late time evolution. It is worth emphasising that this is a mechanism that generates inhomogenous cosmological models starting from an initially homogeneous one. We can see these inhomogeneities appearing explicitly by solving the equation in the relevant regimes.

We can obtain an analytical solution inside and outside the screening radius (equivalently, before or after the screening time). For that, we first obtain the energy conservation equation for the differential equation assuming that there is no shell crossing so that the mass inside each shell is constant. Under these assumptions, we can integrate the evolution equation once to give
\be
E(R)=\frac12 \dot{R}^2-\frac{G M}{R}+\mU(R)\,,
\ee
where we have defined
\be\label{eq:UInt}
\mU(R)=\mU_\star-\frac{\beta^2 GM}{r_s}\int_{r/r_s}^{R/r_s}\frac{\dd x}{\sqrt{1+x^4}}.
\ee
This equation can be recast in the more suggestive form of a Friedmann equation
\be
H^2(t,r)\equiv\frac{\dot {a}^2}{a^2}=\frac{8\pi G\rho_\star}{3a^3}+2\frac{k(r)-\widetilde\mU}{a^2}\,,
\label{hubr}
\ee
where we have defined $k(r)=E(r)/r^2$ and $\widetilde\mU = \mU/r^2$. We will assume that there is no spatial curvature so that we can take $E\simeq0$. At early times, all shells are assumed to be well within their screening radius so $r\ll r_s$ and we can take the lower limit in the integral \eqref{eq:UInt} to be zero. On the other hand, since the Born-Infeld correction is assumed to be negligible in that early stage, it is safe to neglect the contribution $\mU_\star$ so the Born-Infeld energy is well approximated by
\be
\widetilde\mU\simeq-\frac{\beta^2 GM}{{r^2}r_s}\int_0^{R/r_s}\frac{\dd x}{\sqrt{1+x^4}}.
\ee
At early times, well before the screening time, we can neglect the Born-Infeld force and integrate the evolution for $R$, which is given by
\bea
R_{in}(t,r)=r\left[1+\frac32H_\star \big(t-t_\star\big)\right]^{2/3}
\eea
corresponding to a dust dominated universe
We recover that at early times the motion is comoving and the scale factor $a=R/r$ is homogeneous. At later times, the Born-Infeld force is turned on and, asymptotically, its contribution to the energy dominates over the gravitational one, so the evolution is simply given by
\be
R_{\rm out}=\sqrt{-2 \mU_\infty(r)} \;t
\ee
where we have defined 
\bea
\nonumber
\mU_\infty(r)&=&-\frac{\beta^2 GM(r)}{r_s(r)}\int_{0}^{\infty}\frac{\dd x}{\sqrt{1+x^4}}\\ &=&-\frac{4 \Gamma^2(5/4)}{\sqrt{\pi}}\frac{\beta^2 GM(r)}{r_s(r)}\,.
\eea
This evolution corresponds to a curvature dominated universe but the spatial curvature is inhomogeneous. However, this solution gives $\dot{R}/{R}=1/t$ which does not depend on $r$. On the other hand, since $\mU_\infty\propto r^{3/2}$, the scale factor develops an inhomogeneity $a\propto r^{-1/4}$.

A reasonably good solution can therefore be constructed by appropriately matching $R_{\rm in}$ and $R_{\rm out}$. This can be done by imposing continuity so the matching time $t_{\rm m}$ can be computed by solving the equation 
\begin{equation}
R_{\rm in}(t_{\rm m})=R_{\rm out}(t_{\rm m})
\end{equation}
and the shells evolution can be well-approximated by
\begin{eqnarray}
R(t,r)\simeq\left\{
\begin{array}{c}
r\left[1+\frac32H_\star \big(t-t_\star\big)\right]^{2/3}\quad t<t_{\rm m}\\
\\
\sqrt{-2 \mU_\infty(r)} \;t\qquad\qquad\quad t>t_{\rm m}
\end{array}\right.
\label{eq:Ranal}
\end{eqnarray}
The time $t_{\rm m}$ is parameterically of the order of the time $t_{\rm s}$ at which the corresponding shell crosses its screening radius $R(t_{\rm s},r)=\rs(r)$. We find, however, that doing the matching at $t_{\rm m}$, and not $t_{\rm s}$, by imposing continuity for our simple analytical estimates (\ref{eq:Ranal}) gives a more accurate description of the exact numerical solutions.

From the solution for the shells evolution,  we can compute the velocity by using the energy conservation equation so that 
\begin{equation}
v(t,r)=\sqrt{\frac{2 GM(R)}{R}-2\mathcal{U}(R)}.
\label{eq:vcont}
\end{equation}
Alternatively, we could have simply taken $v=\dot{R}$ from \eqref{eq:Ranal}, but using the exact relation from the energy conservation (\ref{eq:vcont}) gives more accurate results.

As we will see below, these analytical solutions give a fairly good approximation for most of the evolution and, in particular, they  capture well the behaviour in the regions where the shell is sufficiently far from the screening radius crossing.  Unfortunately, they are not precise enough to capture the evolution fully near the crossing and, therefore, the profile of the local Hubble factor $H_{\rm obs}$  for  these relevant and interesting scales. However, although the obtained analytical approximate solutions do not  reproduce the exact local Hubble profile quantitatively, it does describe the qualitative behaviour of the profile and it is also accurate quantitatively in the asymptotic regions so that it allows to have a good understanding of the mechanisms at work.

Now that we have introduced the model under consideration and showed how it naturally gives rise to an inhomogeneous cosmological evolution, we will proceed to constructing the corresponding Hubble diagram for this scenario.

\vspace{-0.1cm}
\section{The Hubble diagram}\label{sec:hubble}

\begin{figure}[!t]
\includegraphics[width=0.5\textwidth]{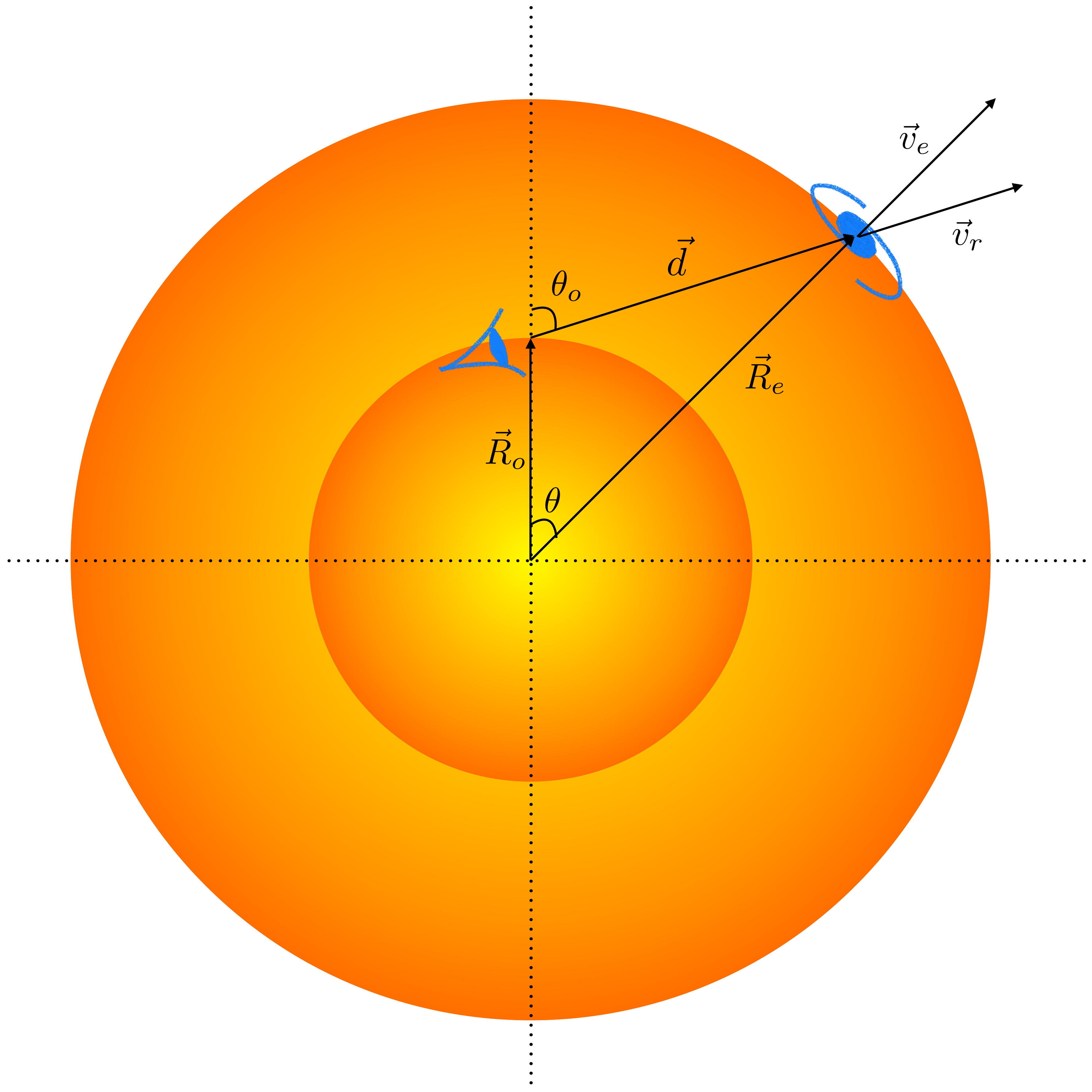}
\caption{In this Figure we show the configuration used to construct the Hubble diagram. The observer is at $\vec R_o$ and the emitter at $\vec R_e$. The angle $\theta$ will play an important role in the Hubble diagram as seen by the observer, contrary to what happens in isotropic and homogeneous cosmological models.}
\label{Fig:Hubble}
\end{figure}

\begin{figure*}[!t]
\includegraphics[width=0.49\textwidth]{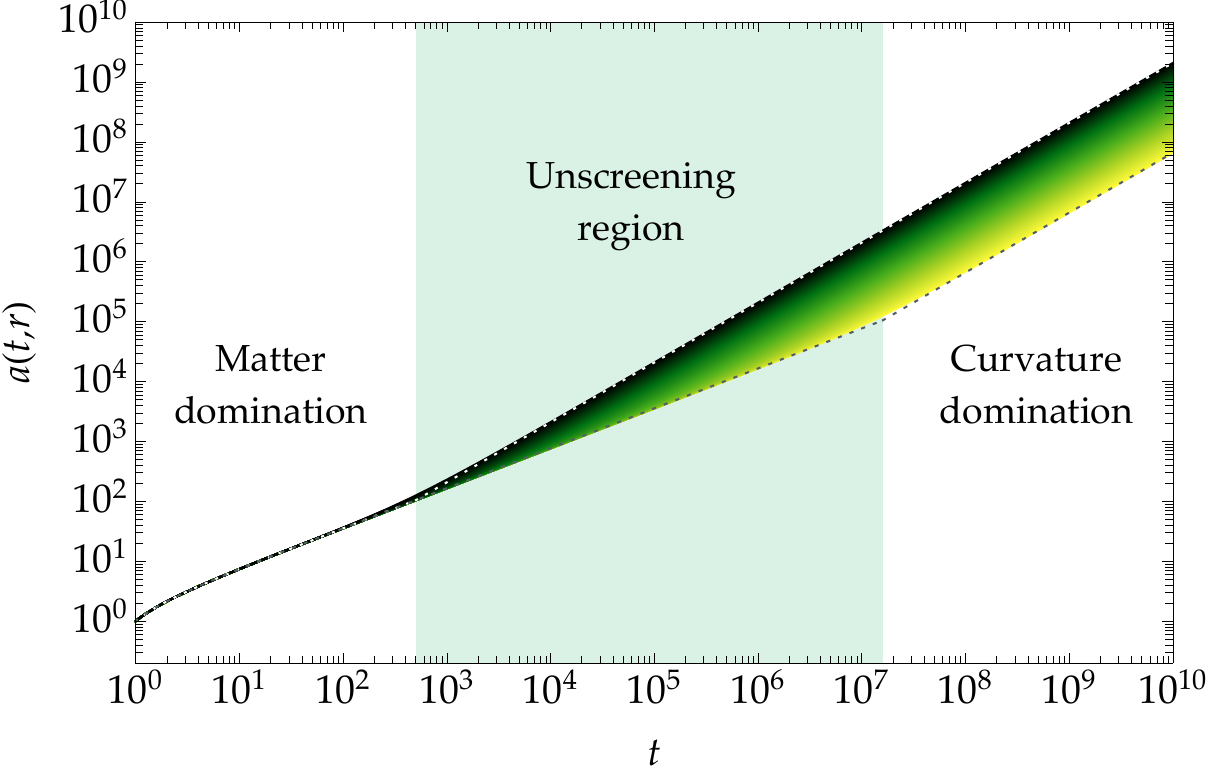}
\includegraphics[width=0.49\textwidth]{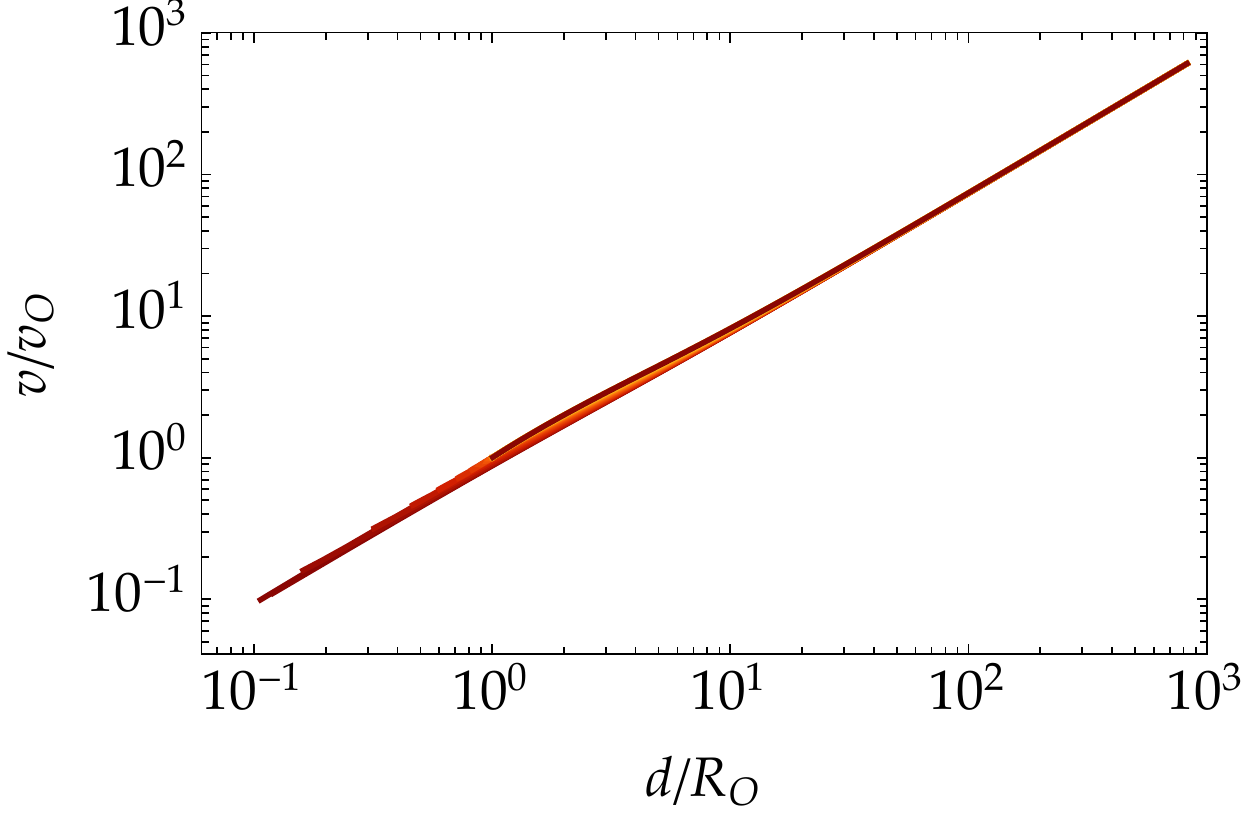}
\caption{In this Figure we show the evolution for the spherical shells obtained numerically. We have used  an ensemble of 100 shells, time units with $H_\star=\sqrt 2$ so that $\frac{4\pi G}{3}\rho_\star=1$. We normalise the size of the shells to the innermost one and our choice of units assigns a unit mass to that shell. In these units, we choose $\Lambdae=10^{-3}$ and $\beta^2=0.9$ that corresponds to having $\as\simeq 77.5$ for the innermost shell and a 10\% reduction of the effective Newton's constant in the unscreened region. The left panel clearly shows how the scale factor develops an inhomogeneous profile as the shells exit their screening radii (light green region). Asymptotically, when all the shells have crossed their respective screening radius, the evolution is again comoving according to a curvature dominated universe, but the generated inhomogeneities in the the scale factor remain. The dotted lines correspond to the approximate analytical solutions of the innermost and outermost shells. We can see how the simple analytical solution \eqref{eq:Ranal} succeed in capturing most of the evolution, including the generated inhomogenious profile, and it only fails (at a few percent level) when the shell crosses its screening radius. In the right panel we construct the corresponding Hubble diagram with the receding velocities and distances computed as explained in the text (normalised to the values of the observer's shell).}
\label{fig:evolutionnoL}
\end{figure*}

\begin{figure}[!t]
\includegraphics[width=0.45\textwidth]{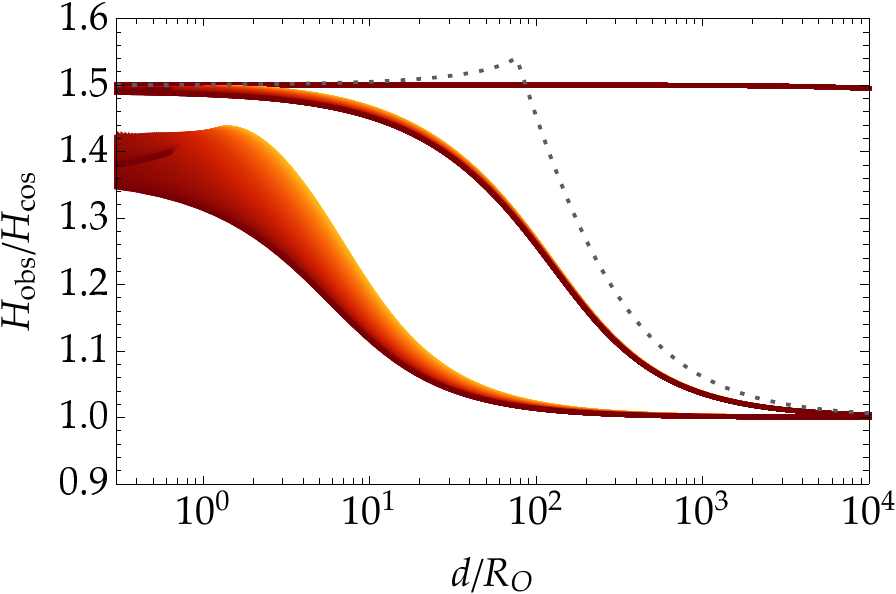}
\caption{We plot the profile of the local Hubble factor normalised to the cosmological Hubble rate corresponding to matter domination ($H_{\rm cos}=\frac{2}{3t}$). We use the same numerical configuration as in Fig. \ref{fig:evolutionnoL} and assume the observer is at the 10th shell. From bottom to top we show the profile at the beginning of the unscreening region ($t=5\times 10^3$), well inside the unscreening region ($t=10^5$) and the curvature-dominated asymptotic regime ($t=10^9$). We can see how the local Hubble rate develops a non-trivial angular dependence when the inner shells start exiting. The dotted line shows the observed Hubble factor computed with the analytical solution \eqref{eq:Ranal} for $t=10^5$. As explained in the text, we see that the asymptotic regions are well captured by the analytical expression, but it fails to reproduce the profile in the region where the shells are close to their screening radius.}
\label{fig:HubblenoL}
\end{figure}

\begin{figure*}[!t]
\includegraphics[width=0.49\textwidth]{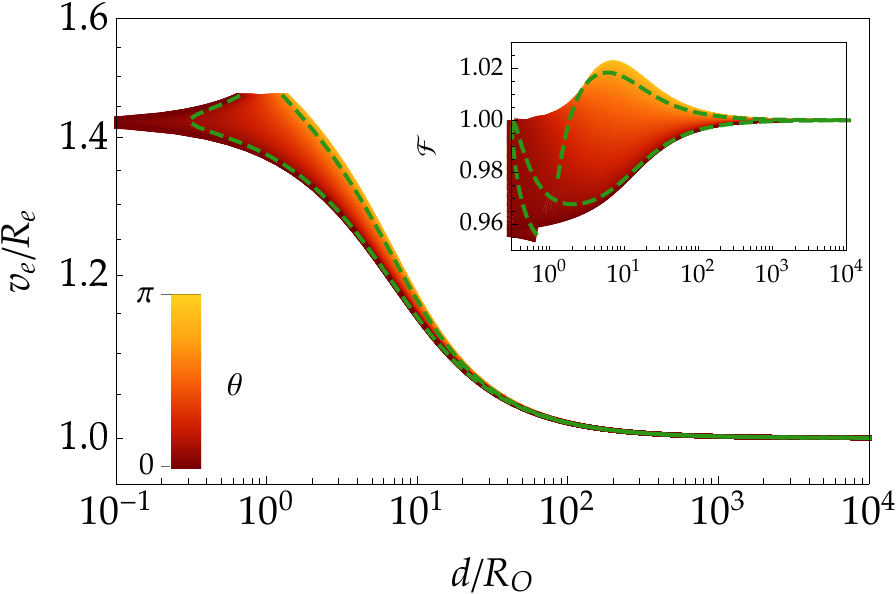}
\includegraphics[width=0.49\textwidth]{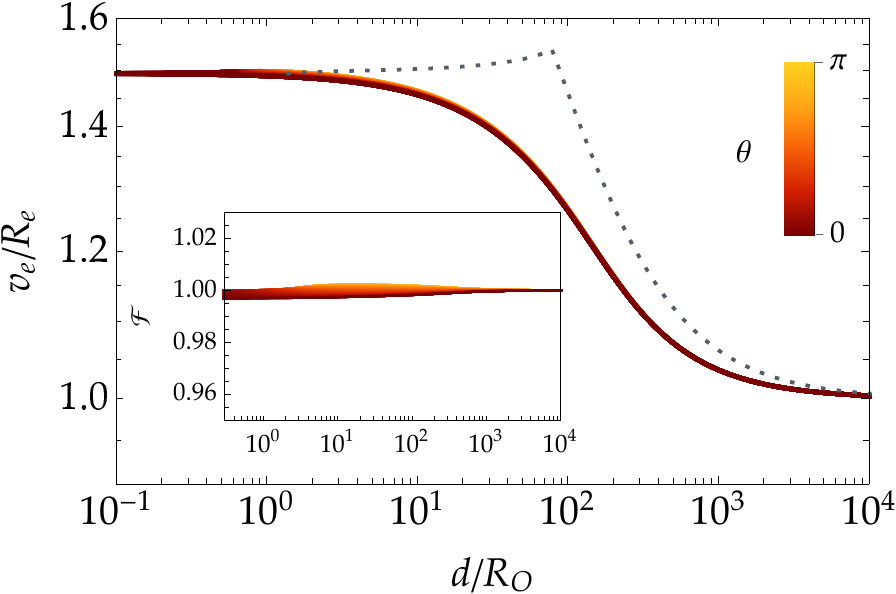}
\caption{In the left panel we show the profile of the emitter's Hubble rate $v_e/R_e$ at a time when the observer's shell just exited its screening radius ($t=5\times 10^3$) when the angular dependence becomes most pronounced (the numerical configuration is the same as in the previous figures). We can clearly see the angular dependence induced by the form factor (shown in the inset). The green dashed lines correspond to the angles $\theta=\pi/10$ and $\theta=3\pi/4$ and we see the double valued character of the form factor for distances $d<R_o$ as explained in the main text. In the right panel we show the same at a later time ($t=10^5$) so the observer is well outside its screening radius. We can see how the angular dependence is much smaller, the form factor approaches 1 and, therefore, the observed Hubble factor is only sensitive to the transition from the curvature domination of the inner shells to the matter domination of the outer shells. The dotted line corresponds to the approximate solution given in Eq. \eqref{eq:Ranal} and we confirm that this simple analytical expression correctly captures the asymptotic regimes, while the zone where the shells are crossing their screening radii is much more poorly described.}
\label{fig:vede}
\end{figure*}
\vspace{2cm}

\vspace{-1.75cm}
Let us now use our Newtonian approach for the cosmology with the screened dark electric force to obtain the Hubble diagram for this model as seen by some observer living in this universe. For that, we will discretize the initially homogeneous distribution of matter by considering an ensemble of spherical and concentric shells that evolve according to Eq. \eqref{eq:eqR} that we solve with the numerical method explained in \cite{Jimenez:2020bgw}. The numerical solutions will also allow us to corroborate our analytical estimates above.

In figure~\ref{Fig:Hubble} we show the configuration from which we are constructing the Hubble diagram. The observer is located on a given shell with radius $R_o(t)$ with the north pole of our spherical coordinate system in the direction of the observer so the angle $\theta=0$ is defined as in figure~\ref{Fig:Hubble}. The observer vector position is then given by $\vec{R}_o=R_o(t)(0,0,1)$ and its velocity will be $\dot{\vec{R}}_o(t)=\dot{R}_o(t)(0,0,1)$. The Hubble diagram is constructed by computing the receding velocity of some emitting source located at $\vec{R}_e=R_e(t)(\sin\theta\cos\varphi,\sin\theta\sin\varphi,\cos\theta)$ with respect to the observer, that is given by
\begin{equation}
    v_r=\big(\vec{v}_e-\vec{v}_o)\cdot\frac{\vec{R}_e-\vec{R}_o}{\vert\vec{R}_e-\vec{R}_o\vert}\,.
    \label{eq:v_r}
\end{equation}
By symmetry, this velocity can only depend on the azimuthal angle $\theta$ and  is independent of the polar angle $\varphi$.
Hence, the observational Hubble factor $H_{\rm obs}$,  corresponding to the recession velocity of the emitter with respect to the observer, can then be expressed as
\be\label{eq:Hubble_factor1}
H_{\rm obs} \equiv\frac{v_r}{\vert\vec{R}_e-\vec{R}_o\vert}=\frac{v_e}{R_e}\frac{1+\frac{v_o R_o}{v_e R_e}-\left(\frac{R_0}{R_e}+\frac{v_o}{v_e}\right)\cos\theta}{1+\left(\frac{R_o}{R_e}\right)^2-2\frac{R_0}{R_e}\cos\theta}\,.
\ee

In the case of comoving motion, where the evolution of the shells can be factorised as ${R_i}(t)=a(t)g(r_i)$, we have ${v}_r=\vert\dot{\vec R}_e(t)-\dot{\vec R}_o(t) \vert=\frac{\dot{a}}{a} \vert \vec R_e(t)-\vec R_o(t)\vert $ and we recover the usual Hubble law $v_r=H_{\rm obs}(t) d$ with a $\theta-$ independent observed Hubble factor. Of course, this just reflects the conservation of homogeneity and isotropy in a comoving evolution. In our scenario, the scale associated to the screening mechanism leads to a violation of the comoving evolution as the shells cross their respective $r_{s,i}$ so that the Hubble diagram relating the velocity $v_r$ to the distance $d= \vert \vec R_e -\vec R_o\vert $ shows a discrepancy from a pure linear relationship and a dependence on the azimuthal angle $\theta$. We will explicitly show this behaviour from our numerical solutions, but let us first give some analytical insight to gain some understanding of the mechanisms at play.

Notice that the distances $R_o$, $R_e$ and $d$ can be related via the geometric relation (this is independent of any cosmology)\begin{equation} 
d^2 = R_e^2 + R_o^2 -2 R_e R_o \cos\theta\,.
\end{equation}
Hence, we can solve for $R_e$ as a function of the other quantities 
\be
\frac{R_e}{R_o}=  \cos\theta \pm \left( \frac{d^2}{R_o^2}-\sin^2\theta\right)^{1/2}\,.
\label{ddd}
\ee
As can be seen, when $d\ge R_0$ the square root is always well-defined and there is only one solution corresponding to the plus sign in the previous expression (so that $R_e(\theta=0)=R_o+d$). 
Similarly when $d\le R_o$, there are two solutions and the angles are restricted to
\be
\vert \sin \theta \vert \le \frac{d}{R_o}\,.
\ee
As a function of $d/R_o$, $R_e/R_o$ is single-valued when $d\ge R_o$ and double-valued when $d\le R_o$.
The Hubble factor \eqref{eq:Hubble_factor1} can be written in a more transparent way by introducing the parameter
\be 
\alpha= \frac{v_o}{v_e}\frac{R_e}{R_o}\equiv \frac{H(t,r_o)}{H(t,r_e)}\,,
\ee
which is the ratio of the cosmological Hubble rates (\ref{hubr}) associated to  the shells to which the observer and the emitter belong, all evaluated at time $t$. 
Using this parameter we obtain that 
\begin{eqnarray}
&&H_{\rm obs}=
\frac{v_e}{R_e} {\cal F} =
\label{He}\\\nonumber
&&=\frac{v_e}{R_e}\left[1+\frac{\alpha-1}{2}\left(1+\left(\frac{R_o}{d}\right)^2-\left(\frac{R_o}{d}\cos\theta \pm\gamma\right)^2\right)\right]\,,
\end{eqnarray}
where
\begin{equation}
    \gamma= \sqrt{1-\left(\frac{R_o}{d}\right)^2\sin^2\theta}\,,
\end{equation}
and where we have introduced the form factor ${\cal F}$ that can be expressed in terms of the observer's angle $\theta_o$ (see figure~\ref{Fig:Hubble}) as
\be
{\cal F}= 1+(\alpha-1) \cos \theta_o \frac{R_o}{d}\,.
\ee
Notice that, at large distance ($d\gg R_o$),  $H_{\rm obs}$ coincides with the cosmological rate of the emitter {and the angular dependence is suppressed regardless of the observer's position}. 
It is worth stressing that so far we have not made any approximations, so that equation for the observed Hubble rate is so far exact. In Fig. 
\ref{fig:vede} we show the dependence of both $v_e/R_e$ and the form factor ${\cal F}$ as a function of $d/R_o$.

At very early times when the evolution is comoving we have that $v_e/v_o=R_e/R_o=r_e/r_o$ so $\alpha=1$ and  we recover the cosmological  result $H_{\rm obs}=\dot{R}_e/R_e=2/(3t)$, which does not depend on the angle nor the distance. At very late times when both the emitter and the observer have crossed their screening radii, they evolve as $R_{e,o}=C r_{e,o}^{3/2}t$, with $C$ a universal constant. In this case, there is no  strict comoving motion  as the scale factors depend on the shells via $r_{e,o}$, see figure \ref{fig:evolutionnoL} where the asymptotic slope can be seen and the dependence on the shells is also clear. However, we have that $v_e/v_o=R_e/R_o=(r_e/r_o)^{3/2}$ implying that $\alpha=1$  leading to the Hubble factor $H_{\rm obs}=v_e/R_e=1/t$ that is again independent of the angle
and corresponds to the cosmological one. 

One of the most interesting regime can be obtained when there is  a difference in the observed Hubble factor as measured locally and at large (cosmological) distances. This can be realised for instance  if the observer has exited its screening radius, but the emitter is still screened. In this case we find that
$ v_o/v_e= (H_e t)^{-1} R_o/R_e$, and using $H_e t \simeq 2/3$ this leads to $\alpha= 3/2$ and therefore we have
\be
H_{\rm obs}=\frac{v_e}{R_e}\left(1+ \cos \theta_o \frac{R_o}{2d}\right) .
\ee
When $d\gg R_o$, this becomes as $\theta_o\simeq \theta$
\be
H_{\rm obs}=H(t,r_e)\left(1+ \cos \theta \frac{R_o}{ 2d}\right) .
\ee
where $H(t,r_e)\equiv \frac{2}{3t}$. As a result, the correction to the observational Hubble rate  is scattered according to the different values of $\theta$. These results are illustrated in  Fig. \ref{fig:HubblenoL}.

\begin{figure*}
\includegraphics[width=0.49\textwidth]{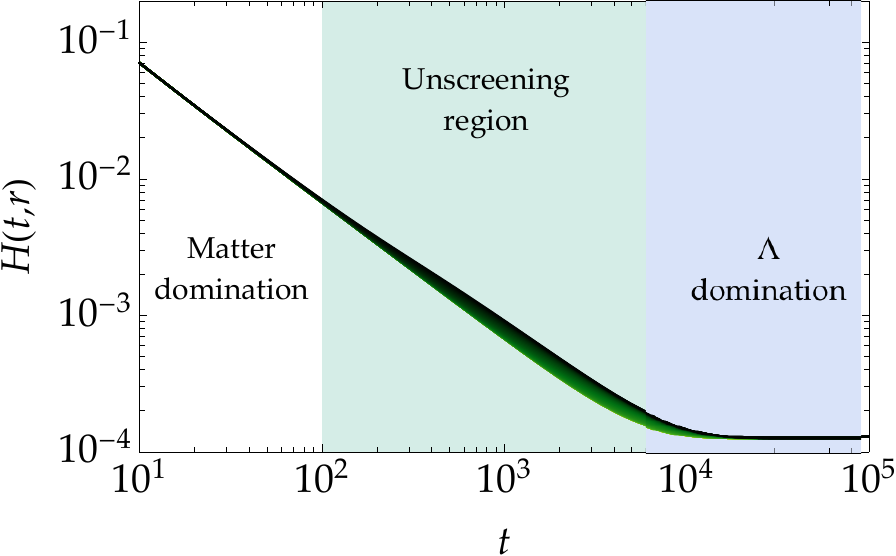}
\includegraphics[width=0.49\textwidth]{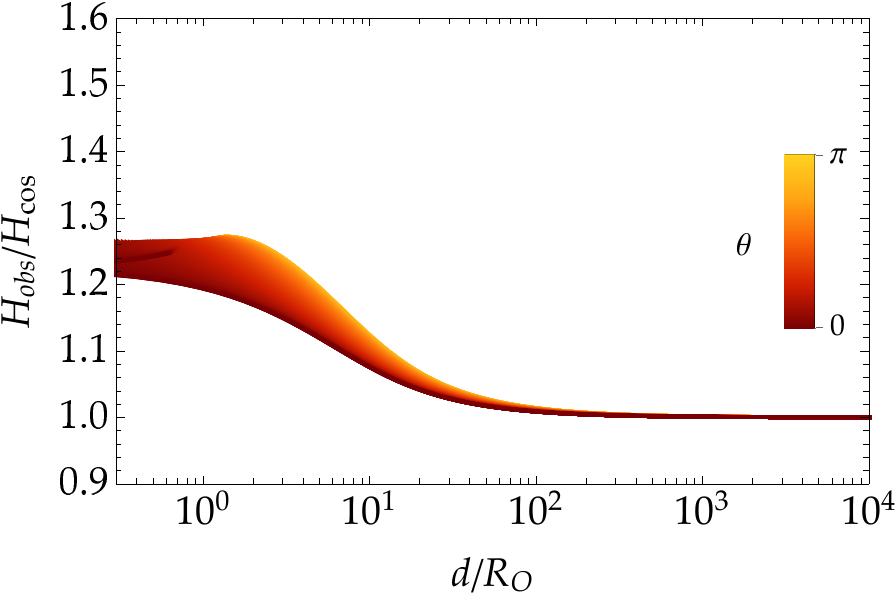}
\caption{In this Figure we show the effect of adding a cosmological constant with energy density $\rho_\Lambda=10^3 \Lambdae^4$ and all the other parameters and initial conditions as in the previous figures. We can see how the cosmological constant prevents the generation of further inhomogeneities once it dominates and also contributes to lowering the difference between the cosmological and the local Hubble factors. {The right panel uses the same conditions as in Fig. 3, at $t= 5\times 10^3$.}}
\label{fig:HubbleL}
\end{figure*}
\vspace{2cm}

\vspace{-1.75cm}
In order to construct the observed Hubble diagrams as defined by Eq.~\eqref{eq:Hubble_factor1} we have numerically solved the equations for 100 shells with initial radii ranging from  1 to $10^6$ in units of the innermost shell with a uniform logarithmic distribution. The initial conditions are chosen to have a vanishing spatial curvature initially, comoving motion and so that all the shells are inside their respective screening radii at the initial time. More explicitly, the initial (purely radial) velocities of the shells are set as $v_i(t_\star)=H_\star r_i$ with $H_\star=\frac{8\pi G \rho_\star}{3}$. 

In the left panel of Fig.~\ref{fig:evolutionnoL} we show the evolution of the scale factor where we can clearly see the three different stages as explained above. The initial evolution is comoving (i.e., the scale factor is homogeneous). As the shells exit their screening radius they abandon the dust dominated evolution and start evolving according to curvature domination. The crossing of the screening radius occurs in a hierarchically manner so the inner shells join the curvature dominated evolution at earlier times. When the largest shell exits its screening radius, the universe exhibits again a comoving evolution but now with an inhomogeneous scale factor. In the right panel of Fig.~\ref{fig:evolutionnoL} we show the Hubble diagram (evaluated at a given time) for this universe, where we can see an angular dependence on small scales. The properties of the Hubble diagram become more apparent in Fig.~\ref{fig:HubblenoL} where we plot the observed Hubble factor normalised to that of a the pure matter dominated evolution in a universe without the screened extra electric force. We can see how very distant objects give rise to the expected cosmological Hubble factor, while nearby objects would give rise to a modified Hubble diagram with two main effects. Firstly, the spatial curvature produced by the electric force when the shells are unscreened give rise to a higher value of the local observed Hubble rate. Secondly, the non-comoving evolution for such shells also introduce a scatter due to the angular dependence. This angular dependence prominently occurs at times when the observer is close to its screening scale. As the observer's shell becomes larger and larger than its screening radius, the form factor approaches  $\mathcal{F}\simeq 1$ for all angles and $v_e/R_e$ is mostly sensitive to the matter-curvature transition (see Fig. \ref{fig:vede}).

So far we have neglected the effect of dark energy in our discussion. However, its inclusion does not change significantly the overall picture. It is immediate to introduce the effects of a cosmological constant in the dynamics so the shells now evolve according to
\be
\ddot{R}=-\frac{GM(R)}{R^2}\left[1-\frac{\beta^2}{\sqrt{1+\left(\frac{\rs(R)}{R}\right)^4}}\right]+\frac{16\pi G}{3}\rho_\Lambda R\,,
\label{eq:eqRL}
\ee
where $\rho_\Lambda$ is the energy density associated to the cosmological constant. In appendix \ref{sec:equations} and (\ref{newt}) we explain how this term arises from the relativistic theory. We show the evolution when the cosmological constant is included in Fig.~\ref{fig:HubbleL}. In the left panel we depict the cosmological Hubble factor, i.e., $H=\frac{\dot{a}}{a}$ where we can see how the cosmological constant governs the late-time evolution, replacing the asymptotically curvature dominated phase of the pure dust case considered above. There remains, however, a transient curvature dominated phase that lasts longer for the inner shells and which may even never be reached by the outer most shells. In any case, it is interesting to note that the inhomogeneous evolution generated by the screening mechanism is still operative in the transient unscreening region. In the right panel of Fig.~\ref{fig:HubbleL} we plot the normalised observed Hubble diagram equivalent to Fig. \ref{fig:HubblenoL} evaluated at $t=5\times 10^3$. This confirms how the cosmological constant lowers the difference between the cosmological and the local Hubble factor. This is so because once the cosmological constant takes over the dynamics, the evolution of all shells is driven by its presence  and the effects of the screening mechanism are attenuated.

\section{Discussion and conclusions}\label{sec:conc}

In this paper we have assumed that dark matter is charged under a new electromagnetic interaction and that this {\it dark} electrodynamics is non-linear. A very well-known example of non-linear electrodynamics is provided by the Born-Infeld theory where the electric field of a point-charge is finite at the centre of the charge distribution. This is an example of screening of the dark electromagnetic interaction which plays a crucial role in our model. Indeed, outside a radius called the {\it screening radius}, the dark point-charge is not screened whereas inside the screening radius the dark electromagnetic interaction is suppressed. We use this phenomenon in a cosmological context. 
In particular, we have explored the phenomenological consequences of having a dark matter sector charged under a dark non-linear electromagnetic interaction on the Hubble diagram. As mentioned above we have used the paradigmatic example of the Born-Infeld theory for its singular properties, but our results are valid for general non-linear electromagnetism as long as there is no shell crossing.\footnote{While for Born-Infeld the absence of shell crossing seems to be guaranteed by its remarkable properties such as the absence of shock waves despite being a non-linear theory, it is not a general feature within the class of non-linear electromagnetisms.}

Initially, all the relevant scales are well inside the screening radius and the dynamics is oblivious to the presence of the charges. Hence, the cosmological evolution is that of a homogeneous, matter dominated universe with all the shells expanding with the same Hubble rate. When the inner shells begin exiting their screening radii they start to feel the repulsive force and inhomogeneities start developing. During this unscreening stage both the scale factor $a$ and the Hubble rate are inhomogeneous. Finally, in the asymptotic, negative curvature dominated regime, where the repulsive force dominates, we end up with an inhomogeneous universe with an homogeneous Hubble rate as can be seen from Fig. \ref{fig:evolutionnoL}. When a cosmological constant $\Lambda$ is added to the picture, its effect is to prevent the generation of further inhomogeneities once it dominates and also to lower the difference between the cosmological and the local Hubble factors.

In order to see in detail the effects of this model on the Hubble rate, we have solved numerically the evolution of an initially homogeneous and isotropic distribution of dark matter particles organised in spherical shells and we have explicitly constructed the local Hubble diagram. A key outcome of these models is  the dynamical generation of an inhomogeneous cosmological evolution and of a temporary inhomogeneous Hubble diagram. The crucial point is the presence of the screening scale. As we have seen, as long as the observer and the emitter belong to the same cosmological stage (either matter dominated or curvature dominated) they both see the same Hubble rate albeit the scale factor is inhomogeneous in the second case. On the other hand, when the observer just exit its screening horizon the Hubble rate becomes inhomogeneous. Moreover, if the emitter is not too far away in the cosmological regime, the Hubble rate will also show an angular dependence.  
It is worth noticing that this scenario can lead to differences of up to a factor of $3/2$ in the Hubble rates of very far objects as compared to nearby ones, even in the regime when the angular dependence becomes negligible. 

Before concluding let us briefly comment on the fact that the inhomogeneous nature of the cosmological evolution in the presence of a dark and screened electromagnetic interaction should have further dynamical consequences. For example, large scales structures dynamics could 
develop anomalous flows due to the repulsion between galaxy clusters (their halos). Structure formation should also be hampered as clumping should appear to be more difficult to realise. All this could be studied for instance by large N-body simulations taking into account the new electromagnetic interaction. A dedicated study of the spherical collapse model of structure formation should also be rewarding. Finally, the inhomogeneous properties of the local Hubble diagram that we have uncovered in this paper should be compared to present data, in particular one of their main theoretical features, i.e., the scatter of the Hubble diagram due to its anisotropy, should certainly provide strong constrains on the model. We will come back to all these points in future works. 

\section*{Acknowledgments}

JBJ and DB acknowledges support from the ``Atracci\'on  del  Talento  Científico'' en Salamanca programme and from project PGC2018-096038-B-I00 by Spanish Ministerio de Ciencia, Innovaci\'on y Universidades. DB acknowledges support from  Junta de Castilla y Le\'on and Fondo Europeo de desarrollo regional through the project SA0096P20. This project has received funding /support from the European Union’s Horizon 2020 research and innovation programme under the Marie Skłodowska -Curie grant agreement No 860881-HIDDeN.

\appendix

\section{Relativistic field equations}
\label{sec:equations}

In this appendix we discuss how the equations solved within the Newtonian approach can be derived from a set of relativistic covariant equations. Here, we will not restrict to the specific case of Born-Infeld Lagrangian, but we will consider a generic non-linear function for the dark electromagnetic sector. We assume General Relativity to describe gravitational interaction and we take as matter content a non-linear electromagnetic field, a charged dust component and a cosmological constant term $\Lambda$.
Let us consider the following action 
\begin{equation}
    S=\int d^4x\left[\sqrt{-g}\mK(Y,Z) + J_q^\mu A_\mu\right]+S_{\rm fluid}[g,n]\,,
\end{equation}
where $\mK(Y,Z)$ is a function of both $Y\equiv -F_{\mu\nu}F^{\mu\nu}/4$ and $Z\equiv -F_{\mu\nu}\tilde F^{\mu\nu}/4$ and $J_q^\mu$ is the charge density current. The action for the fluids is given by \cite{Schutz:1977df,Brown:1992kc}
\begin{equation}
    S_{\rm fluid} = \int d^4 x\left[\sqrt{-g} \rho(n) + J^\mu \partial_\mu l\right]\,,
\end{equation}
where $l$ is a set of Lagrangian multiplier and $J^\mu$ is the mass density current.  

Varying the above action with respect to the dynamical variables gives the following set of equations
\begin{eqnarray}
&&G_{\mu\nu} +\Lambda g_{\mu\nu}= 8 \pi G \sum_i T^{(i)}_{\mu\nu}\,,
\\\label{eq:F_cov}
&&\nabla_\nu\left(\mathcal{K}_Y F^{\mu\nu}+\mathcal{K}_Z \tilde{F}^{\mu\nu}\right)=  \frac{J_q^\mu}{\sqrt{-g}}\,, 
\\
 &&\nabla_\mu T^\mu_{(d)}{}_\nu = F_{\nu\mu} \frac{J_q^\mu}{\sqrt{-g}}\,,
\end{eqnarray}
where we have labelled  the dark fluid $(d)$ and the dark electromagnetic component $(e)$. The conserved current is given by
\begin{equation}\label{eq:current_covariant}
    J_q^\mu = q J_d^\mu=q\sqrt{-g}nu^\mu
    \,,
\end{equation}
with the constant $q$ expressing the charge per particle and with $u_d^\mu$ the four velocity of the dark fluid. The energy momentum tensor for the fluid is
\begin{equation}
    T^\mu_{(d)}{}_\nu=(\rho_d+\pred)u_d^{\mu}u_{d\nu} +\pred\delta^{\mu}{}_\nu\,,
\end{equation}
while that of the dark electromagnetic field is
\begin{equation}
   T_{\mu\nu}^{(e)} = \mathcal{K}_Y(Y,Z) F_{\mu}{}^\alpha F_{\nu\alpha} + g_{\mu\nu}\left(\mathcal{K}(Y,Z)-Z\mK_Z\right)\,.
\end{equation}  
In what follows we will assume $\mK(Y,Z)$ possesses a $\mathbb{Z}_2$ symmetry with respect to $Z$. This implies that it is not contradictory to set $\vec B=0$. Hence, the last term between brackets in equation \eqref{eq:F_cov} drops out \cite{Jimenez:2020bgw}.

\subsection{Lema\^itre model}\label{sec:lemaitre}

The presence of a net charge forbids the existence of homogeneous cosmological solutions.\footnote{A possible way of preserving homogeneity is to give up gauge invariance by adding a mass term for the Maxwell field \cite{1979ApJ...227....1B}.} Hence, one needs to resort to inhomogeneous models in order to discuss the cosmological evolution. The proper choice for the case at hand is the Lema\^itre metric \cite{Lemaitre:1933gd,Misner:1964je} (See \cite{Bolejko:2011jc} for a review on inhomogeneous cosmological models)
\begin{equation}
\label{eq:metric}
    ds^2 = - e^{A(r,t)}dt^2 + e^{B(r,t)} dr^2 + R^2(r,t)d\Omega^2\,.
\end{equation}
The Einstein equations derived from this metric can be cast in the following form
\begin{eqnarray}
\label{eq:EFElemaitre_Mprime}
    \mathcal M'&=&4 \pi G(\rho_d +\rho_e) R^2 R'\,,\\
    \label{eq:EFElemaitre_Mdot}
    \dot{\mathcal M} & =& -4 \pi G (p_d+p_e) R^2\dot R\,,\\
    \label{eq:EFElemaitre_constr}
    A'&=&2\frac{\dot R'}{\dot R}-\frac{R'\dot B}{\dot R}\,,
\end{eqnarray}
where the dot and the prime denote derivation with respect to coordinate time $t$ and the radial coordinate $r$ respectively and where we have introduced the mass \cite{Lemaitre:1933gd,Misner:1964je,Hernandez:1966zia}
\begin{equation}\label{eq:mass}
    2\mathcal M=R\left(1+\dot R^2 e^{-A}-R'^2 e^{-B}\right)-\frac{\Lambda R^3}{3}\,.
\end{equation}
Equation \eqref{eq:EFElemaitre_constr} can be formally integrated to give
\begin{equation}
\label{eq:BofA}
    e^B = \frac{R'^2}{1+2E(r)}e^{-\int A'\frac{\dot R}{R'}dt}\,,
\end{equation}
 where $E(r)$ is an arbitrary time-independent function. The equation defining the mass \eqref{eq:mass} can be recast into an evolution equation for $R$
\begin{equation}
        e^{-A}\frac{\dot R^2}{R^2} = \frac{2 \mathcal \mathcal M}{R^3}+\Lambda- \frac{1-(1+E(r))e^{\int A'\frac{\dot R}{R'}dt}}{R^2}\,.
    \end{equation}

In order to describe the matter content we will assume that there exists a coordinate system $(t,r,\theta,\phi)$ such that the motion of the dark fluid is comoving, i.e., the velocity $u^r=0$. Notice that this does not contradict the fact that there is a Hubble flow determined by $R(r,t)$\footnote{For a discussion of this point see \cite{1970JMP....11.1382C}.}. This is consistent with the initial conditions (screened cosmological evolution) but it further requires the absence of shell-crossing. In what follows we will assume that we do not have to worry about this issue either because the theory itself prevents it (as is the case of Born-Infeld) or because we are considering the dynamics until this occurs.
With this choice, the isotropic but inhomogeneous charged perfect fluid has an energy momentum tensor of the form
\begin{equation}
    T^\mu_{(d)}{}_\nu=(\rho_d+\pred)u_d^{\mu}u_{d\nu} +\pred\delta^{\mu}{}_\nu\,,
\end{equation}
where $u_d^t = e^{-A/2}$ and $u_d^i = 0$ for $i=r,\theta,\phi$, and where $\rho_d$ and $\pred$ depend on both $t$ and $r$. 

Furthermore, thanks to the spherical symmetry of the Lemaître metric, we are allowed to consider $F_{tr}$ as the only non-vanishing component of the dark electromagnetic field strength, i.e., we are considering only a radial electric field.

Hence, the equations for the non-linear electromagnetic field reduce to
\begin{eqnarray}
\frac{d}{dr}\left[R^2 \mathcal{K}_Y F^{tr} e^{(A+B)/2}\right]&=&R^2  q n u^t e^{(A+B)/2}\,,
\end{eqnarray}
which can be integrated to give
\begin{equation}\label{eq:Ftrup_eq}
    \mK_Y F^{tr} = \frac{Q e^{-(A+B)/2}}{4\pi R^2}\,,
\end{equation}
where we have defined the charge enclosed in a certain radius $r$ as
\begin{equation}
    Q(r) = 4\pi \int dr R^2 q n e^{B/2}\,.
\end{equation}
Let us stress again that this results holds if there is no shell crossing. If this were not the case there would be a net flux of charged particles through a given surface and hence one could not set to zero the current $J^r$. 

The equations for the dark fluid read
\begin{eqnarray}
\label{eq:dotepsd}
& 2\dot \rho_d &+ \left(\dot B+4 \frac{\dot R}{R}\right)(\rho_d+ \pred)=0\,,\\
\label{eq:predprime}
   & \pred'+ &\frac{\rho_d+\pred}{2}A'= -q n_d F_{tr}e^{-A/2}\,.
\end{eqnarray}
Now, using \eqref{eq:EFElemaitre_constr} to express $\dot B$ and using \eqref{eq:predprime} we can rewrite the continuity equation as
\begin{equation}
    \dot \rho_d +\left(2\frac{\dot R}{R}+\frac{\dot R'}{R'}\right)(\rho_d+\pred)=+\frac{\dot R}{R'}\left(\pred'+ q n_d F_{tr}e^{-A/2}\right)\,.
    \end{equation}

From now on we specialise to the case of charged dust, i.e., we set $p_d=0$ \cite{AIHPA_1973__18_2_137_0,Misra:1974gb}. Then, the equations governing the dynamics reduce to
\begin{eqnarray}\label{eq:Aprime_F}
    &&\frac{A'}{2}= \frac{q n_d}{\rho_d}F_{tr}e^{-A/2}\,,\\
    \label{eq:rho_F}
   &  \dot \rho_d & +\left(2\frac{\dot R}{R}+\frac{\dot R'}{R'}\right)\rho_d=\frac{\dot R}{R'}\left(q n_d F_{tr}e^{-A/2}\right)\,,\\
   \label{eq:Hubble_R_rel}
     &&   e^{-A}\frac{\dot R^2}{R^2} = \frac{2 \mathcal M}{R^3}+\Lambda -\frac{1-(1+2E(r))e^{2\mathcal{U}}} {R^2}\,,
    \end{eqnarray}
where $F_{tr}$ is the solution to the non-linear equation \eqref{eq:Ftrup_eq} and where we have defined
\begin{equation}
    \mathcal{U}=\int \frac{A'}{2}\frac{\dot R}{R'}dt\,.
\end{equation}
Eq.~\eqref{eq:Aprime_F} makes clear how the inhomogeneities in the metric are sourced by the electric field while Eq.~\eqref{eq:rho_F} implies that an initially homogeneous dark fluid distribution will eventually become inhomogeneous. 

\subsection{The Newtonian limit}

In the Newtonian limit, $A$ is a small Newtonian potential implying that the exponent in Eq.~\eqref{eq:BofA} can be neglected. Hence, the evolution equation \eqref{eq:Hubble_R_rel} reduces in this limit to
\begin{equation}
        \frac{\dot R^2}{R^2} = \frac{2 \mathcal M}{R^3}+{\Lambda} + \frac{2E(r)}{R^2} + \frac{2{\cal U}}{R^2}\,.
\end{equation}
The Newtonian limit of the integral $\mathcal{U}$ can be conveniently written as
\begin{equation}
\mathcal{U}=-\frac{4\pi G\rho_\star R^3_\star}{3r_s}\beta^2\int_{R_\star/r_s}^{R/r_s}\frac{dx}{x^2\mK_Y(x)}\,,
\end{equation}
where we have defined the constant $\beta = \sqrt{2}M_{\rm Pl}q/m$, the initial radius of the shell $R_\star=R(t_\star,r)=r$ and we have introduced the  scale $r_s$  characterising the transition between the screened and unscreened regimes that is specified once a choice for the function $\mK$ is made.
    Furthermore, since we are considering dust ($p_d=0$) the Einstein equations \eqref{eq:EFElemaitre_Mdot} imply that the mass of the dust component is conserved.\footnote{This is true in the absence of shell crossing \cite{Jimenez:2020bgw}, which is the case for the Born-Infeld model we are considering in the main text.} { {Moreover, we can neglect the electromagnetic contribution to the mass $\mathcal M= {\cal M}_d +{\cal M}_e$. In fact, combining the solution to the Maxwell equation with that of the mass one can see that the electromagnetic density of a shell is of order $\mathcal M_e/R^3\simeq \beta^2 G^2 M^2(R)/R^4$ to be compared against a matter density of order $\mathcal M_d/R^3=G M(R)/R^3$. The electromagnetic contribution is small provided $\beta^2 \sim 1$, i.e., the dark interaction has gravitational coupling strength and we use for the Newtonian potential $\Phi_N(R)\simeq G_N M(R)/R\ll 1$.\footnote{Another way to see why the electromagnetic mass does not contribute to the Newtonian limit is as a $1$-loop effect. In fact, while the mass density directly sources the Einstein equations the charge does so via the Maxwell equation. Since the electromagnetic energy density is proportional to the square of the field strength the charge enters the Einstein equations as an integrated effect proportional to the square of the coupling that in our case is $\mathcal O(\beta^2 G)$. }   Cosmologically and even on smaller galactic and solar scales, Newtonian potentials are at most of order $10^{-4}$ implying that for $\beta={\cal O}(1)$ the Newtonian approximation is valid.} Hence, we can identify 
\be
\mathcal M=G M_d= \frac{4\pi G}{3}\rho_\star R_\star^3\,,
\ee
with $\rho_\star$ being the initially uniform distribution of charged matter and in the second term we have connected the mass that appears in the Einstein equations with  that of \eqref{eq:mass_Newt_d}.
Hence, in the Newtonian limit, we get that the evolution equation for the radius of a spherical shell is
\begin{equation}
    \frac{\dot R^2}{R^2} = \frac{8\pi G\rho_\star}{3}\left(\frac{R_\star}{R}\right)^3 +\Lambda + \frac{2E(r)}{R^2}+\frac{2\mathcal{U}}{R^2}\,.
    \label{newt}
\end{equation}
To get a closer connection with cosmology, it is convenient to foliate space-time with
\be
R(t,r)= r a(t, r)
\ee
where initially all the shells at $t=t_\star$ are such that $a(t_\star,r)=1$, i.e., the initial radial foliation allows one to have a Lagrangian description of the evolution of the shell. In the regime we are considering, the radial gradient of the scale factor is small, i.e., $\partial_r \ln a(t,r)\ll r^{-1}$.
Hence,
\be \label{eq:IB}
{\cal U}= \frac{4\pi G \beta^2 r^2\rho_\star}{3} \int_1^a \frac{da}{a^2 {\cal K}_Y (a/a_s)}\,,
\ee
where we have denoted by
\be 
a_s(r)= \frac{r_s(r)}{r}
\ee
the scale factor at crossing of the screening radius for a shell of initial radius $r$. 
In particular ${\cal K}_{Y}(x)\to \infty $ for $x\ll 1$ and ${\cal K}_{Y}(x)\to 1 $ for $x\gg 1$.
The evolution equation reduces to the modified Friedmann equation
\begin{equation}\label{eq:hubble}
\frac{\dot a^2}{a^2} = \frac{8\pi G \rho_\star}{3a^3}+\Lambda +2 \frac{k(r)}{a^2} + 2\frac{{\cal U}}{r^2 a^2}\,.
\end{equation}
where $k(r)=E(r)/r^2$ and $\mathcal U$ is given in \eqref{eq:IB}.
We can easily derive Newton's equation by taking the time derivative of $\dot a^2$, namely
\be
\ddot a = -\frac{4\pi G \rho_\star}{3a^2} + \Lambda a +\frac{4\pi G  \rho_\star}{3} \frac{\beta^2}{a^2 {\cal K}_Y (a/a_s)}.
\ee
The first term corresponds to the gravitational interaction, the second one to the cosmological constant and the last one is the dark force action on a shell of radius $R$ due to the electric field. This equation matches Eq. \eqref{eq:eqa} that was derived within the Newtonian approach and when the function $\mK$ is the Born-Infeld one.

\section{Dark scalar electrodynamics}
\label{app:DSE}
In this appendix, we present a simple dark scalar electrodynamics model able to encompass the model discussed in this work. Let's consider the action
\begin{eqnarray}\label{eq:DSE_action}\nonumber
 S=\int & d^4x & \sqrt{-g}\left[\frac{M^2_P}{2}R+\mathcal{K}(Y,Z)\right.\\
 &-&\left. \frac{g^{\mu\nu}}{2}D_{(\mu}\chi(D_{\nu)}\chi)^*-m^2\chi\chi^*\right]
\end{eqnarray}
where an asterisks means complex conjugate and 
where we have defined
\begin{equation}
D_\mu\chi=(\nabla_\mu-iqA_\mu)\chi\,.
\end{equation}
Notice that this Lagrangian is manifestly gauge invariant and the associated conserved current is
\begin{equation}
   J^\mu=-iq\left(\chi^*\partial^\mu\chi-\chi\partial^\mu\chi^*\right)-2q^2\chi\chi^*A^\mu\,.
\end{equation}
In order to make a connection with the fluid description adopted in this work it is convenient to use the Madelung representation for the complex scalar field $\chi$  
\begin{equation}
    \chi = \frac{1}{m}\sqrt{\frac{\rho}{2}}e^{i\theta}\,,
 \end{equation}
where the bare scalar field mass $m$ has been introduced so that $\rho$ has the appropriate dimensions for a density. In terms of these new variables the action \eqref{eq:DSE_action} becomes
\begin{eqnarray}
\nonumber
    S&=&\int d^4x\sqrt{-g}\left[\frac{M_P^2R}{2}-\frac{\partial_\mu\sqrt{\rho}\partial^\mu\sqrt{\rho}}{2m^2}-\frac{(\sqrt{\rho})^2}{2m^2}\partial_\mu\theta\partial^\mu\theta-\right.\\ \nonumber
    &&\left.\frac{(\sqrt{\rho})^2}{2 m}-\frac{g^{\mu\nu}}{2}\left(J_\mu A_\nu+J_\nu A_\mu\right)-\frac{q^2}{m^2}(\sqrt{\rho})^2A^2+\mK\right]\,,\\
\end{eqnarray}
with
\begin{equation}
    J_\mu = \frac{q}{m} (\sqrt{\rho})^2\left(\frac{1}{m}\partial_\mu\theta-\frac{q}{m}A_\mu\right)\,.
\end{equation}
The Klein-Gordon equation becomes a system of two equation for $\rho$ and $\theta$:
\begin{eqnarray}
&
\Box\theta +\frac{\partial_\mu\rho}{\rho}\partial^\mu\theta-qA^\mu\frac{\partial_\mu\rho}{\rho}-q\nabla_\mu A^\mu=0\,,\\   
&
-\partial_\mu\theta\partial^\mu\theta = m^2 - \frac{\Box\sqrt{\rho}}{\sqrt{\rho}}-2qA_\mu\partial^\mu\theta+q A^2\,.
\end{eqnarray}

Let us now introduce the vector $u^\mu\equiv\partial^\mu\theta/m$ and take the covariant (uncharged) derivative of the second equation in the above system. Exploiting the fact that the ``velocity'' $u^\mu$ is the gradient of the scalar field, we get
\begin{equation}
   \left(u^\mu-\frac{q}{m}A^\mu\right))\nabla_\mu u_\nu = \frac{1}{2m^2}\nabla_\nu\left(\frac{\Box\sqrt{\rho}}{\sqrt{\rho}}\right)+\frac{q}{m}u^\mu\nabla_\nu A_\mu\,.
\end{equation}
subject to the constraint
\begin{equation}
    -u^\mu u_\mu= 1 - \frac{1}{m^2}\frac{\Box\sqrt{\rho}}{\sqrt{\rho}}-2\frac{q}{m}A_\mu u^\mu+\frac{q^2}{m^2}A^2\,.
\end{equation}
Let us now define a new vector $v^\mu = u^\mu -q/m A^\mu$. With this shift in the vector $u^\mu$, the equations read
\begin{eqnarray}
&&\nabla_\mu(\rho v^\mu)=0\,,\\
&& v^\mu\nabla_\mu v_\nu =\frac{1}{2m^2}\nabla_\nu\left(\frac{\Box\sqrt{\rho}}{\sqrt{\rho}}\right)-\frac{q}{m}v^\mu F_{\mu\nu}\,.
\end{eqnarray}
The constraint on the norm of the vector $v^\mu$ takes the form
\begin{equation}
    -v_\mu v^\mu = 1-\frac{1}{m^2}\frac{\Box\sqrt{\rho}}{\sqrt{\rho}}\,.
\end{equation}

Notice that, despite the appearances, this is not the system of equations for a perfect fluid. Firstly, the velocity vector $v$ is not normalized to unity but rather satisfies the constraint above. Secondly, as we shall see below, the energy momentum tensor is not that of a perfect fluid. In terms of the new vector $v^\mu$ the current takes the simple form
\begin{equation}
    J^\mu = -\frac{q}{m}\rho v^\mu\,.
\end{equation}
In terms of these variables, the  stress energy tensor reads
\begin{eqnarray}\nonumber
    T_{\mu\nu} &=& \rho v_\mu v_\nu +g_{\mu\nu}\left(-\frac{1}{2m^2}\partial_\alpha\sqrt{\rho}\,\partial^\alpha\sqrt{\rho}-\frac{\rho}{2}(v_\alpha v^\alpha+1)\right)\\
    \nonumber
    &+&\frac{1}{m^2}\partial_\mu\sqrt{\rho}\partial_\nu\sqrt{\rho}+\mK_Y F_{\mu}{}^\alpha F_{\nu\alpha}+g_{\mu\nu}\left(\mK-Z\mK_Z\right)\,.\\
\end{eqnarray}

The energy momentum tensor cannot be reduced to that of a perfect fluid, as it is clear from the complex nature of the scalar field. However, in the limit in which the derivatives of the density are small compared to the mass scale set by the parameter $m$, it is easy to see that we are actually dealing with a pressureless perfect fluid \cite{Bettoni:2013zma}, hence recovering the relativistic dynamics described in the appendix \ref{sec:equations}.

\bibliography{fluid}

\end{document}